\documentclass[a4paper,11pt]{article}
\usepackage[T2A]{fontenc}
\usepackage[cp1251]{inputenc}
\usepackage{amsmath,amsthm,amssymb}
\usepackage{graphics}
\usepackage{graphicx}
\begin{document}
\title{Study of the itinerant electron magnetism of Fe-based superconductors by the proximity effect}
\author{Yu. N. Chiang$^{1}$, M. O. Dzuba$^{1, 3}$, O. G. Shevchenko$^{1}$ and A. N. Vasiliev$^2$\\
\emph{$^{1}$B. I. Verkin Institute for Low Temperature Physics and Engineering,}\\
\emph{National Academy of Sciences of Ukraine, 47 Lenin ave., Kharkov 61103, Ukraine.}\\
\emph{$^2$ Lomonosov Moscow State University,}\\
\emph{GSP-1, Leninskie Gorky, Moscow, 119991, Russian Federation.}\\
\emph{$^{3}$International Laboratory of High Magnetic Fields and Low Temperatures,}\\
\emph{53-421 Wroclaw, Poland.} \\
\emph{E-mail:chiang@ilt.kharkov.ua}}
\date{}
\maketitle
\begin{abstract}
We used the proximity effect as a tool to achieve an ideal ("barrier-free") NS boundary for
quantitative evaluation of transport phenomena that accompany converting dissipative current into
supercurrent in NS systems with unconventional superconductors – single-crystal chalcogenide FeSe and
granulated pnictide LaO(F)FeAs. Using features (limitations) of Andreev reflection in the NS systems
with dispersion of the electron spin subbands, we revealed direct evidence for spin-polarized nature
of transport and the absence of residual magnetization in iron-based superconductors in the normal
state: In heterocontacts with single-crystal FeSe and granular LaO(F)FeAs, we detected a
spin-dependent contribution to the efficiency of the Andreev reflection associated with the spin
accumulation at the NS boundary, and a hysteresis of conductivity of FeSe in the ground state in low
external magnetic fields. Based on our findings, we conclude that in iron-based superconductors, the
itinerant electron magnetism is predominant, magnetism of iron atoms being localized.
\end{abstract}
\section{Introduction}
Along with the discovery of superconductivity in multi-component compounds with elements having
significant local magnetic moment (Fe, Ni) [1], the detection of the same phenomenon in two-component
compounds with the same elements proved to be an important discovery [2, 3]. Thus, a number of
multi-component superconducting compounds of different composition, including iron-based, is closed by
a compound directly adjacent to the family of single-element conventional superconductors. In this
regard, there is no doubt that the appearance of superconductivity in multi-element compounds with
delocalized electrons is closely related to the reduced symmetry of the crystal, in particular, such
as the symmetry of "layered" type. This symmetry is characteristic of structures in a large family of
compounds, containing a wide range of rare earths, pniktogens, chalcogens, and transition elements Mn,
Fe, Co, Ni, Cu, and Ru. It leads to the anisotropy of the electronic and magnetic properties
accompanied by an increased electron density of states in the layers with quasi-two-dimensional
(anisotropic) Fermi surface and by an increased role of electron-electron interaction. Anisotropy of
the properties seems to be that feature under which condition in the same material, the magnetic
interactions coexist with the interactions that generate superconducting pairing of the excitations in
the electron subsystem of layered superconductors of complex composition, including iron-based ones.

By now, the notion of crystal structures of layered superconductors and the nature of coupling in them
is sufficiently developed and experimentally established, while their magnetic and electronic
structures in the ground state are still a subject of debate and active research. In this regard, it
is of considerable interest to compare electronic properties of layered iron superconductors which
share the crystal structure of PbO type (\emph{P}4/\emph{nmm}) that predetermines related
quasi-two-dimensional structures of the electronic bands with nesting [4 - 6]; relevant examples are
the binary phase $\alpha$ - FeSe and oxypnictide LaO(F)FeAs.

Theoretical collective efforts using local density of states approximation (see, eg, [4, 7, 8]) lead
to the conclusion that the mechanism of superconductivity in iron-based pnictides and chalcogenides is
likely to have nothing to do with the electron-phonon mechanism, even when the value of the critical
temperature, $T_{c}$, does not extend beyond the McMillan criterion [9], based on the values of the
coupling constants and phonon dispersion characteristic of the electron-phonon pairing concept. This
conclusion is sufficiently proved, despite the fact that the calculations "from the first principles"
by the density functional method can give certain ambiguity in the definition of density of states
and, as a consequence, of the band structure at the Fermi level [10 - 12]. In any case, the main
argument in favor of this conclusion is as follows: The superconductivity in the presence of magnetic
elements is observed in a wide range of compounds with layered crystalline structure of the same type
and in a wide range of critical temperatures both satisfying and not satisfying (exceeding) the
electron-phonon criterion.

One of the common methods of studying the electronic properties of superconductors is, as known, the
investigation of the Andreev conductance, $G_{if}$, of NS interfaces, either artificially created
(heterosystems) or naturally produced in a homogeneous material, in the form of wide or narrow
channels [13]. As the number of open Andreev levels ($N_{\bot}$) is directly proportional to the cross
section of the interface, $\mathcal{A}$ ($N_{\bot} \sim \mathcal{A}/\lambda^{2}_{\rm F},\ {\rm where}\
\lambda_{\rm F}$ is the Fermi wavelength of an electron), it is clear that to study Andreev
conductance is generally preferable to use wide channels. In this case, the only inconvenience for an
experimenter, in the absence of artificial barriers on the interface (\emph{z} = 0 where \emph{z} is a
parameter characterizing the energy barrier strength), is essentially low resistance of the channels.
Intrinsic barrier height and a corresponding resistance which we will call an own resistance of the
interface $R_{if}^{\rm NN}$ (known as the "Sharvin resistance" of restrictions [14]), are inversely
proportional to the cross section of the interface: $R_{if}^{\rm NN} = (G_{if})^{-1} = (p_{\rm
F}/e^{2}n)/\mathcal{A} \equiv 3 [2N(0)e^{2}v_{\rm F}]^{-1}/ \mathcal{A}$ (here, $e, p_{\rm F},~v_{\rm
F},\ n,\ {\rm and}\ N(0)$ are the charge, the Fermi momentum and velocity, the concentration and
density of states per spin for free electrons, respectively).
\begin{figure}[htb]
\begin {center}
\includegraphics[width=10cm]{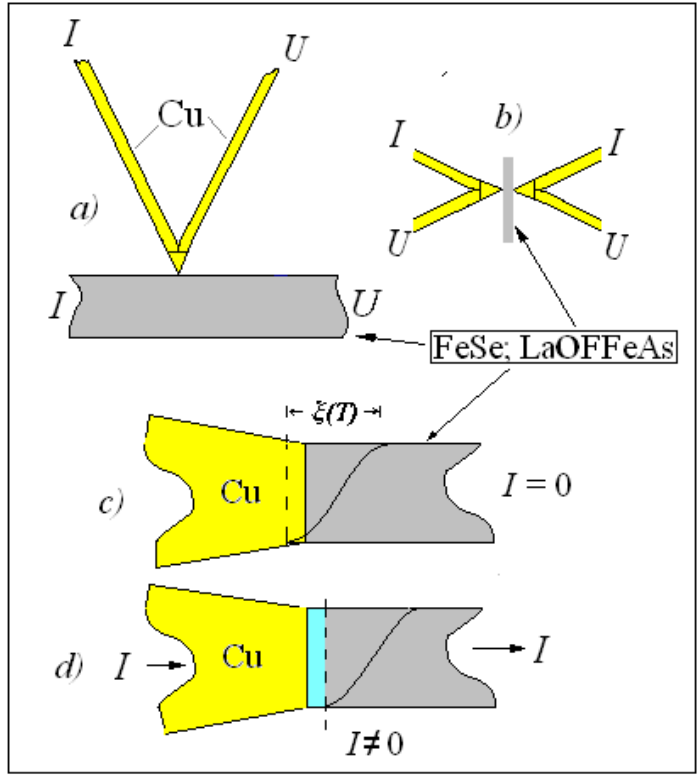}\\
\caption{(\emph{a}) and (\emph{b}) illustrate the ways to measure IVC of the point-contact samples;
(\emph{c}) displays dispersion of the order parameter induced by the proximity effect near the NS
boundary (dashed line) in the absence of transport current \emph{I}; (\emph{d}) depicts a change in
the position of NS boundary (dashed line) after the current is turned on. The region of the
superconductor passed to the normal state is highlighted. See text for details.} \label{1}
\end{center}
\end{figure}
For substances with the conductivity of conventional metals, $R_{if}^{\rm NN} \approx 5 \cdot
10^{-11}/\mathcal{A}\ [\Omega]$ if $\mathcal{A}$ is expressed in cm$^{2}$. It follows that the
restrictions (interfaces) with a diameter of the order of $1\ \mu$m cannot have an own resistance
exceeding $\sim 10^{-3}\ \Omega$. Greater values often demonstrated by real point contacts indicate
that, along with the resistive contribution from the interfaces $R_{if}^{\rm NN}$, dissipative
contributions exist from extraneous inclusions, such as parts of a probe (eg, the tip of the scanning
probe) or defective or oxide barriers. In most cases, this is due to the fact that a "four-probe"
method of measuring the current-voltage characteristics (IVC) when applying to point contacts, appears
to be essentially two-contact one (see Fig. 1 \emph{a, b}). That is why an area of incomparably
greater length than ballistic one is forced to be measured; therefore, not only own contribution from
the interface in the ballistic approximation is gauged [15]. In other words, the real contacts of
point geometry, in general, cannot be considered ballistic in case of electrical measurements [16].

Physics of such contacts includes several mechanisms controlling the value of the system conductance
$G_{pc}^{\rm NS}$ in the NS state of the interface. After the system switches from the NN to NS state,
total resistance of a real point contact, $R_{pc}^{\rm NS}$, contains at least the following additive
contributions: (\emph{i}) a dissipative contribution from the N - side of the interface of the total
length of $L^{\rm N}$ (by this we mean an overall contribution from a part of the tip, from an oxide
layer, and from the layer which thickness measures alike the coherence length where the scattering
cross section by impurities doubles under Andreev retroreflection [17]); (\emph {ii}) an own
contribution from the interface $R_{if}^{\rm NN}$ with the weight determined by the efficiency of the
Andreev reflection which is a function of the energy parameters of the system (electron energy and the
energy gap of a superconductor [15]); (\emph{iii}) a dissipative contribution from the part of a
superconductor related to the dispersion of the order parameter at the NS interface [18] due to the
proximity effect. At the NN $\rightarrow$ NS transition, the contribution (\emph{i}) generally
decreases the conductance of the contact and the contribution (\emph {ii}) increases it. Previously,
we have shown [16] that in a barrier-free non-ballistic contact, the contribution (\emph{i}), in
general, should prevail over the contribution (\emph{ii}) within the energy range $k_{\rm B}T;\ eU \ll
(L^{\rm N} / l_{el}^{\rm N}) k_{\rm B}T_{c}$; here, $T,\ U,\ l_{el}^{\rm N},\ {\rm and}\ T_{c}$ are
the temperature, bias voltage, electron mean free path in the N – side, and the critical temperature
of the superconductor, respectively.

Among these contributions, the contribution (\emph{iii}) is the least known, especially that aspect of
the proximity effect which is associated with the ability to generate a perfect NS boundary. Indeed,
due to the dispersion, the order parameter at the NS boundary changes from 1 to 0 over the spatial
range of the order of the Ginsburg-Landau characteristic length scale $\xi(T)$ (Fig. 1 \emph{c}). This
means that the NS boundary can be moved by a macroscopic magnetic field of the transport current, no
matter how small it is, deep into the superconductor, as shown in the Figure, provided that that field
can suppress the superconductivity in the S-side of the contact. Thus, at finite values of the
transport current, the NS boundary can be an ideal interface between the two parts of the same
superconducting material which are in different states - normal and superconducting.

In this paper, we used this feature of the proximity effect for studying the nature of the magnetism
of the ground state in new iron-based superconductors, chalcogenide FeSe and oxypnictide
La[O$_{1-x}$F$_{x}$]FeAs. One might hope that the absence of extraneous inclusions at the NS boundary,
often with uncontrolled characteristics that reduce the informativeness of Andreev reflection
phenomenon, allows to judge with certainty about the presence or absence of the dispersion of the spin
subbands in the normal ground state of a superconductor. Thus, we will be able to understand whether
magnetism (and, indirectly, superconductivity) of iron-based superconductors is mainly itinerant and
long-range phenomenon or localized and short-range one. Since the contributions (\emph{i}) and
(\emph{iii}) are directly proportional to the thickness of the respective layers, their weight should
be more noticeable as a total length of the NS sample approaches these thicknesses. Hence, we are led
to maximum possible "shortening" of this length, ie, to a point-contact geometry of the samples and to
the schemes for measuring CVC shown in Fig. 1 and corresponding, as explained above, with a
two-contact measurement design. Here, we present the results of the research of Andreev conductance of
non-ballistic point-contact NS heterostructures with relatively wide interfaces (of the area
$\mathcal{A} \sim 10^{-4}\ {\rm cm}^{2}$). We study the systems Cu/FeSe and
Cu/La[O$_{1-x}$F$_{x}$]FeAs.

\begin{figure}[htb]
\begin{center}
\includegraphics[width=10cm]{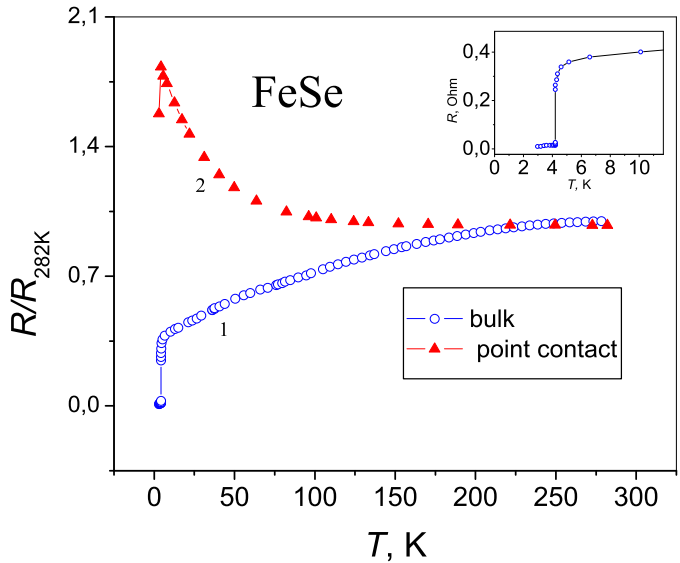}\\
\caption{Resistance of FeSe samples as a function of carriers energy set by the temperature \emph{T}:
1 - bulk samples measured in the geometry of non-concurrent probes (Inset shows the scaled-up region
of the superconducting transition); 2 – point-contact samples measured in the geometry of the
concurrent probes depicted in Fig. 1 \emph{a, b}.}\label{2}
\end{center}
\end{figure}
\section {Experiment}
Starting materials of the superconductors used for preparing hybrid samples with point-contact
geometry have a different structure in line with the technology of their preparation. Pnictide
La[O$_{0.85}$F$_{0.1}$]FeAs was obtained by solid-phase synthesis, such as described in Ref. [19], and
had a polycrystalline structure. X-ray diffraction and spectroscopic studies have shown the presence
of this phase in an amount of not less than 97 \%. Iron chalcogenide was made as a single crystal. To
obtain it we used the technology of crystallization from solution in the melt KCl/AlCl$_{3}$ at
constant temperature gradient 5 K over the range 47 K. Typical sizes of the single crystals are $1.5
\times 1.5 \times 0.5$ mm$^{3}$. X-ray studies carried out on an automatic single-crystal
diffractometer "Xcalibur-3" (Oxf. Diffr. Ltd) show that the crystals of both materials belong to the
tetragonal space group \emph{P}4/\emph{nmm} of PbO type and have lattice parameters $a, b = 3.765$
\AA; $c = 5.518$ \AA~ (these almost repeat the data from Ref. [20]) for the basic binary $\alpha$ -
phase FeSe and $a, b = 4.035$ \AA; $c = 8.729$ \AA~ for fluorine-doped oxypnictide LaO(F)FeAs.

Point-contact samples were produced by mechanically clamping method. As a result they proved to be
very high-resistive (up to several Ohms) due to preferential contribution to their resistance from
oxide layers, the temperature behavior of the resistance being generally of a semiconductor type. CVC
measurements were conducted using stabilized dc power supply. Its output resistance was lower than
that of the samples, which required maintaining the same selected measurement mode, namely the
constant-current regime within all current, temperature, and field ranges during the experiment.
Represented here are the results obtained in this mode of electrical measurements. Generally speaking,
the constant-voltage regime is more informative, though rarely used. The latter, however, is not
technically accessible in all the intervals of control parameters in a single measurement cycle.
\section{Results and discussion}
\begin{figure}[htb]
\begin{center}
\includegraphics[width=10cm]{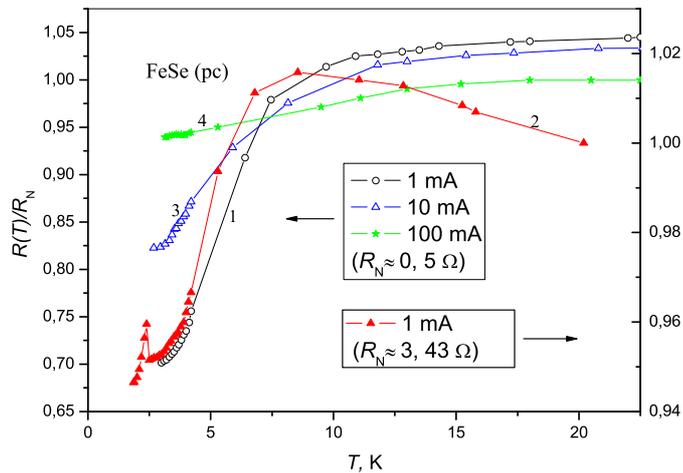}\\
\caption{Normalized temperature dependences of the resistance of point contacts Cu/FeSe measured at
different transport currents.}\label{3}
\end{center}
\end{figure}

In Fig. 2, over a wide temperature range, we demonstrate the difference between the dependencies of
the resistance of the bulk FeSe sample on the carriers energy set by the temperature \emph{T}, derived
from the conventional four-probe measurement geometry, and the same dependencies for point-contact
samples Cu/FeSe with oxide barriers at the interface obtained in the measurement geometry of combined
probes shown in Fig. 1 \emph{a, b}. We see that the difference is qualitative: The character of the
formers corresponds to the metallic behavior while that of the latters is semiconducting. In all the
samples with FeSe, both bulk and hybrid, the superconducting transition was observed in the range of
$(4 \div 5)$ K (see Inset). In general, the above features of the temperature behavior are also
characteristic of both bulk and hybrid NS samples with LaO(F)FeAs which experiences a superconducting
transition at $\sim 26$ K.

Figs. 3 and 4 show typical temperature dependencies of the normalized resistance of point contacts
Cu/FeSe and Cu/LaO(F)FeAs measured at different transport currents. It is seen that, while increasing
measuring current in the studied range $\mathcal{I} = 1 \div 100$ mA, the share of a normal part of
the point-contact system is increasing while that of a superconducting part corresponding to the
change in the contact resistance at the superconducting transition is decreasing. The absolute value
of the superconducting jump in resistance is the same for contacts with different total resistances
but measured at the same current [compare curves 1 ($R_{\rm N}=0.5~\Omega$) and 2 ($R_{\rm
N}=3.4~\Omega$) in Fig. 3]. This indicates that the interface resistance in the samples with pressed
point contacts, in addition to the temperature-dependent semiconductor-type part, contains a
temperature-independent part of the type of residual resistance which does not vanish at $T
\rightarrow 0$.
\begin{figure}
\begin{center}
\includegraphics[width=10cm]{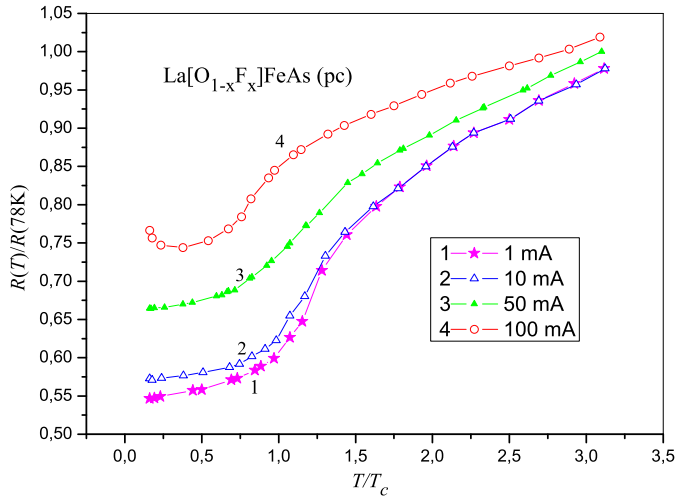}\\
\caption{Normalized temperature dependencies of the resistance of point contacts Cu/LaO(F)FeAs
measured at different transport currents.}\label{4}
\end{center}
\end{figure}

\begin{figure}[htb]
\begin{center}
\includegraphics[width=10cm]{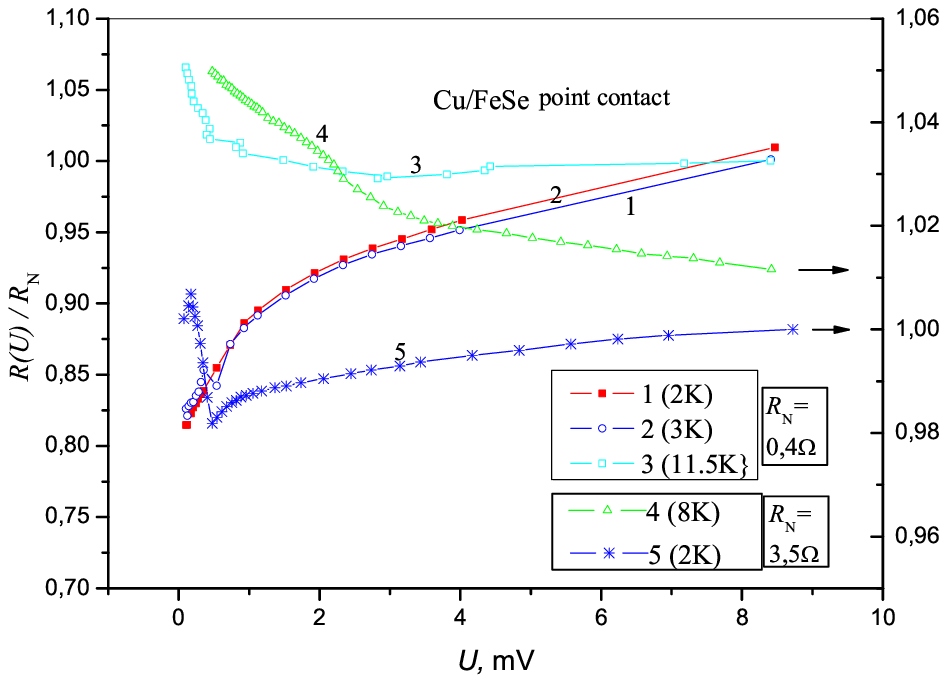}\\
\caption{Normalized resistance of the heterojunction Cu/FeSe measured at different temperatures as a
function of bias voltage \emph{U} at the contact as an addition to the energy $k_{\rm B}T$.}\label{5}
\end{center}
\end{figure}

In Fig. 5, presented are the dependencies of the resistance for the system Cu/FeSe on the bias voltage
\emph{U} at the contact as an addition to the energy $k_{\rm B}T$. Applying that voltage is an
alternative method of controlling energy of the carriers. Curves 1, 2, and 5 correspond to the
temperature range covering the area of the superconducting transition ($T \leq T_{c}$, NS mode) while
curves 3 and 4 were taken in the same interval of bias voltages at temperatures $T > T_{c}$ (NN mode).
It is seen that in comparable energy ranges, the resistance behavior depending on the parameters
\emph{T} (Figs. 2, 3) and \emph{U} (Fig. 5) is qualitatively similar in both NN and NS interface
modes, the values of contact resistance in the NN mode differing by an order of magnitude. From this
and from a comparison between curve 2 in Fig. 2 and curves 3 and 4 in Fig. 5, an important conclusion
follows that in the energy range supporting heterocontacts in NN mode, the semiconductor type of
behavior of the generalized contact conductance most likely is due to the energy-dependent dissipative
contribution from the oxide layer, as the conductivity of FeSe over the entire temperature range above
$T_{c}$ has a metallic behavior (Fig. 2, curve 1) while the resistance of the part of the copper probe
included in the measurement, as small as $\sim 1\ \mu$m in length, cannot exceed a few $\mu \Omega$ at
liquid helium temperatures. The theory predicts [18] that the proximity effect in "dirty" conductors
extends to a depth proportional to the electron mean free path. Hence, at typical thickness of the
"dirty" oxide layers $\simeq 50$\ \AA, the electron mean free path in them is apparently of the same
order. In other words, the residual resistance of contacts in NS state below the superconducting
transition is formed by the intrinsic contribution from the interface ($R_{z}$) which is not
associated either with oxide layers or with the N-side of the interface. As can be seen from Figs. 3,
4, and 5, $R_{z}$ for different contacts amounts to ($70 \div 95$) \% of the normalizing resistance
$R_{\rm N}$ measured in NN mode just before the superconducting transition. Most likely, $R_{z}$
reflects the presence in the heterojunction of a potential barrier of the Schottky type, or the
contribution from surface localized states, or both.

We believe that all these features of the experimental data are directly related to the proximity
effect associated with the dispersion of the order parameter $|\psi|$ which is especially significant
in the range of \emph{T} not too far from $T_{c}$, where its spatial scale defined by the
Ginsburg-Landau coherence length $\xi_{T}\sim (1-T/T_{c})^{-1/2} \xi_{0}$, is quite large. This can be
seen by comparing the magnetic energy \emph{W} of tangential self-magnetic field of the current
$H_{\mathcal{I}}(x=0)=2 \mathcal{I}/r \simeq (10^{-2}\div 2)$ Oe at the penetration depth
$\lambda_{T}$ with the potential of electron pairing $|\psi|=\Delta=\hbar v_{\rm F}/\xi_{0}\sim k_{\rm
B}T_{c}$. The notations are as follows: \emph{r} is the radius of the channel, \emph{x} is the
coordinate measured from the interface on the side of a superconductor occupying a half-space $x>0,\
v_{\rm F}$ and $\xi_{0}$ are the Fermi velocity and the correlation length. The estimation by the
formulae of phenomenological theory [18] leads to the following result:
\begin{equation}\label{1}
W=wV=\frac{H_{\mathcal{I}}^{2}}{8\pi}\mathcal{A}\lambda_{T}\gg \Delta,
\end{equation}
[in the interval $\lambda_{T}$, we replaced the distribution $H_{\mathcal{I}}(x)$ by $H_{\mathcal{I}}$
= const; $\mathcal{A}$ is the interface area ($\geq 10^{-4}\ {\rm cm}^{2}$)]. Thus, for a current of 1
mA, with a typical for London superconductors $\lambda_{T} \sim 0.1\ \mu$m, the energy of the
self-magnetic field of the current amounts to $W \sim 5$ meV, while the value of $\Delta$ is just
$\sim 0.5$ meV. (Note, incidentally, that indicated strength of this inequality for currents greater
than 1 mA is also preserved for Pippard superconductors with $\lambda \sim 0.03\ \mu$m.) Thus, the
examined current interval is suitable for the manifestation of the effect discussed. Moreover, the
fact that the current self-field at $\mathcal{I}=100$ mA can eliminate the manifestation of the
superconducting transition, either almost completely for FeSe (Fig. 3, curve 4) or significantly for
LaO(F)FeAs (Fig. 4, curve 4), means the following. First, the Ginsburg-Landau parameter
$\kappa=\lambda_{T} /\xi_{T}\geq 1$ and hence discussed superconductors, must be characteristic of the
London type-II superconductors and, second, the thickness of the superconducting part of the contacts
is of order of the London penetration depth $\lambda_{\rm L}\simeq 0.2\ \mu$m [18]. It follows that
the length of the measured contact area is of the order of a micron or slightly more - the typical
mesoscopic size which turns out to be non-ballistic due to the presence of the areas with even shorter
elastic mean free paths of electrons and because of the contact geometry characterized by combined
current and potential probes [16].

Inequality (1), of course, overstates the requirements for the value of $H_{\mathcal{I}}$ needed to
suppress superconductivity in the range of $x=\lambda_{T}$, since it implies $H_{\mathcal{I}}$ to be
constant over the whole length of the interval. However, as noted above, the suppression should be
also implemented at lower values of the field at a spatial scale of $\xi_{T}$, due to the dispersion
of the order parameter $|\psi|$ in the proximity effect area. Indeed, the distribution of the magnetic
energy in the superconductor side ($x>0$) depends on the law of magnetic field distribution along the
length of the penetration depth; according to the phenomenological theory [18], it can be written as
\begin{equation}\label{2}
H^{\ast}=H(0)\exp(-x/\lambda).
\end{equation}

\begin{figure}[htb]
\begin{center}
\includegraphics[width = 12cm]{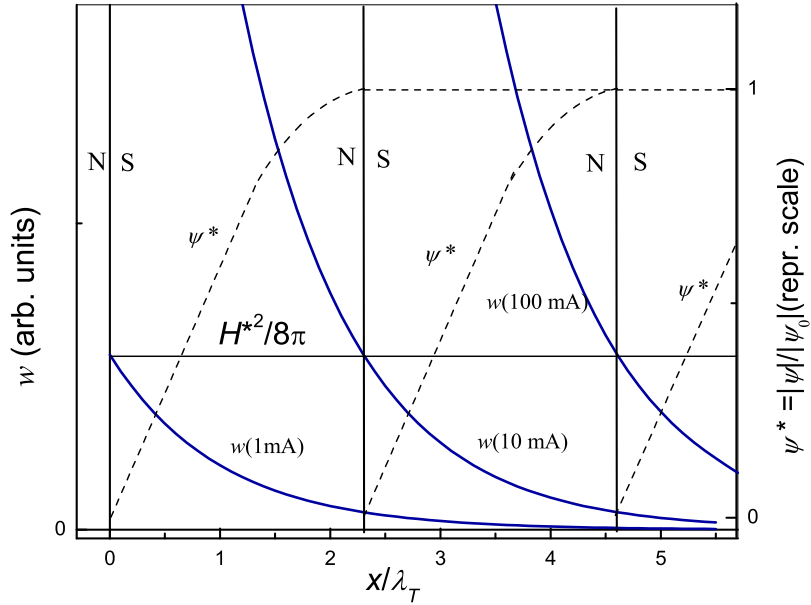}\\
\caption{Displacement of the NS boundary by transport current at the dispersion of the order parameter
caused by the proximity effect.} \label{6}
\end{center}
\end{figure}

Let $H^{\ast}$ is the lowest field, energy density of which $w$ is comparable in value to the reduced,
due to the proximity effect, value of $|\psi|$ normalized to a volume unit. Then, the suppression of
superconductivity at arbitrary $H(0)$ will extend to a depth of \emph{x} depending on $H(0)\equiv
H_{\mathcal{I}}(x=0)$, ie, on the transport current value $\mathcal{I}$, up to that value of \emph{x}
at which one reaches $w=(H^{\ast})^{2}/8\pi$. For larger \emph{x}, ie, for $H<H^{\ast}$, the
superconducting state persists. Thus, the value of $x(H^{\ast})$ sets the position of a defect-free NS
boundary between normal and superconducting parts of the superconductor. From this point, let us
analyze, for example, the data depicted in Fig. 3. From a comparison of Eq. (2) for two values of
$H_{\mathcal{I}}(0)$ it follows
\begin{equation}\label{3}
x_{i}-x_{k} = \lambda_{T} \ln \frac{\mathcal{I}_{i}}{\mathcal{I}_{k}},
\end{equation}
where \emph{i, k} = 1 mA, 10 mA, or 100 mA; $i \neq k$. From the system of pairwise equations we find
\begin{eqnarray}
x_{(1{\rm mA})} & = & 0 \nonumber\\
x_{(10{\rm mA})} & = & \lambda_{T} \ln10\\
x_{(100{\rm mA})} & = & \lambda_{T} \ln100 \nonumber.
\end {eqnarray}
Here, the displacement of the boundary is measured relative to its position at $x_{(1\ {\rm mA})}$
which position, as shown by measurements at $\mathcal{I}<1$ mA (see below), is very close to the
starting point \emph{x} = 0 in the absence of current. Based on Eqs. (4), Fig. 6 explains the physics
of this effect that would be impossible in the absence of the proximity effect in the fields the
maximum value of which, as we have, does not exceed $H(\mathcal{I}=100\ {\rm mA}) \approx 2$ Oe. For
FeSe, for example, that value is more than an order of magnitude lower than that of the first critical
field [21].

The possibility to get a defect-free NS boundary inside a superconductor allows us to interpret with
reasonable certainty the effects associated with converting the dissipative current into supercurrent
through the mechanism of Andreev reflection. On $R(T)/R_{\rm N}$ temperature dependencies measured at
$\mathcal{I} = 1$ mA for Cu/FeSe (Fig. 3, curve 2) and at $\mathcal{I} = 100$ mA for Cu/LaO(F)FeAs
(Fig. 4, curve 4) as well as on $R(U)/R_{\rm N}$ bias-voltage dependencies for Cu/FeSe (Fig. 5, curve
5), gap features are clearly visible. These are positive jumps in the contact resistance at the
transition from the NN to NS regime which value amounts to around 2-3 \% (with resolution better than
0.1 \%) relative to the resistance $R_{\rm N}$ before the superconducting transition. Similar feature
is barely visible on curve 1, Fig. 5, while controlling the energy by bias voltage \emph{U}. It can be
easily understood by noting that under conditions where most of the energy of the electrons is set by
the temperature, an addition $eU$ required for the realization of a gap peculiarity is too small at $T
= 3$ K and for low-resistance sample ($R_{\rm N}\sim 0.5\ \Omega$); besides, in a constant-current
mode (see above), the current itself is small. Indeed, the area of the feature controlled by bias
voltage at 3 K (curve 1, Fig. 5) corresponds to the current of order of 0.04 mA, while that current at
2 K (curve 5, Fig. 5) amounts to around 0.5 mA. Thus, it is of the same order as the current 1 mA
which value corresponds to the discussed jump in resistance controlled by temperature in the contacts
Cu/FeSe with the resistance $R_{\rm N}$ greater by almost an order of magnitude ($\approx 3.5\
\Omega)$. This should mean that the self-field of the current 0.04 mA is too small to invade into a
dispersion region of the order parameter on the FeSe side in NS-interface mode, while at
$\mathcal{I}=1$ mA, the NS boundary is already shifted deep into the superconductor and thus presents
an ideal boundary between normal and superconducting parts of FeSe at which the conversion of the
dissipative current into non-dissipative one takes place. In contacts Cu/LaO(F)FeAs, this specific
feature manifests itself only at the current $\mathcal{I}=100$ mA, probably due to too low value of
$T/T_{c}$ at 4.2 K where $\xi_{T} \approx \xi_{0}$.
\begin{figure}[htb]
\begin{center}
\includegraphics[width = 12cm]{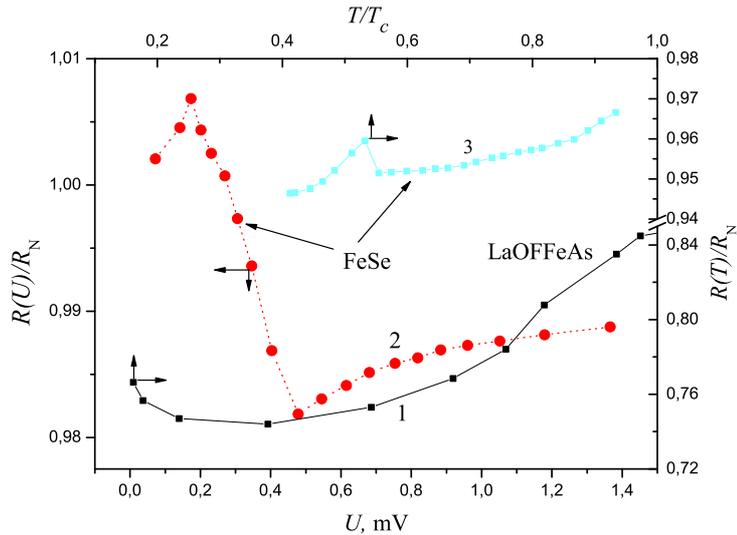}\\
\caption{The effect of spin accumulation - increased resistance of spin-polarized FeSe and LaO(F)FeAs
at the NS interface in the transition from the NN to NS regime.}\label{7}
\end{center}
\end{figure}

In NS systems with non-magnetic superconductors, gap features in the conductance upon Andreev
reflection can occur because of the coherent scattering by impurities on the normal side of the
interface [17, 13] [we recall that we have included the relevant contribution to the resistance from
this effect into the total resistance of the contact N-side and denoted it by (\emph{i}), see Sec. 1].
In systems with magnetic superconductors, a decrease in the conductance of a contact [contribution
(\emph{iii})] at the transition to a sub-gap energy region should also occur as a consequence of the
spin polarization of the current due to the limitation of the Andreev reflection process in the
presence of the dispersion of the spin subbands related by magnetic exchange interaction [22]. As a
result of these limitations, the accumulation of spin should occur at the interface [23, 24]
accompanied by the appearance of a positive addition to the applied bias voltage which, given
constant-current mode, manifests itself as an additive to the total resistance of the contact in the
NS regime. We associate the features observed on the $R(U)$ and $R(T)$ dependencies shown in Figs. 3,
4, and 5 with this additive. From these figures it also follows that the residual resistance of the
contacts in the NS regime, particularly at the current of 1 mA, is by several orders of magnitude
greater than the possible coherent effect of increasing in the resistance in the normal region of the
probe as already noted for the systems in the same situation [16]. This allows us to attribute the
observed features to the manifestation of only the spin accumulation effect. As a direct consequence
of the spin polarization, the effect thus gives an indication of the magnetic characteristics of a
superconductor in the normal ground state which is implemented in a finite region of NS heterojunction
due to the proximity effect.

Fig. 7 shows an enlarged view of these features depending on both control parameters, temperature and
bias voltage, for the contacts with FeSe and on temperature for the contact with LaO(F)FeAs.
Previously, we observed the effect of spin accumulation in the systems "ferromagnet-superconductor"
Fe/In and Ni/In where its value reached 20 \% and 40 \%, with degree of spin polarization of 45 \% and
50 \%, respectively [25]. Recall that the nature of the effect in the NS systems with conventional
ferromagnets is associated with the destruction of the symmetry with respect to spin rotation which
imposes limitations on the probability of Andreev reflection and results from high internal magnetic
field that manifests itself in the spontaneous magnetization. Manifestation of spin accumulation in
unconventional superconductors in normal ground state also points to the absence of such symmetry, but
it cannot be associated with the bulk magnetization, for the latter, in our opinion, is not compatible
with superconductivity at the microscopic level. Thus, once again we get the arguments in favor of
that the magnetism of iron-based superconductors is limited by band magnetism of conduction electrons,
for example, of the type of antiferromagnetic exchange interaction [26].

Knowing the magnitude of the effect of spin accumulation indicating the presence of spin polarization
of the conduction electrons in the normal ground state of FeSe (strictly speaking, of the
superconducting phase of FeSe in the normal state which can amount to as low as 15 \% [26]), we
estimate the polarization factor of the current \emph{P} in studied contacts with single-crystal FeSe.
According to the theory [23, 24], corresponding normalized additive to the resistance $R_{\rm N}$ of
the spin-polarized area due to spin accumulation is of the scale
\begin{eqnarray}\label{5}
\frac{\delta R_{\rm N/S}}{R_{\rm N}} & = & \frac{P^{2}}{1-P^{2}}, ~\\
P & = & (\sigma_{\uparrow}- \sigma_{\downarrow})/\sigma \nonumber; ~
\sigma = \sigma_{\uparrow}+\sigma_{\downarrow};\nonumber\\
R_{\rm N} & = & \lambda_{s}/(\sigma \mathcal{A}).\nonumber
\end{eqnarray}
Here, $\sigma,\ \sigma_{\uparrow},\ \sigma_{\downarrow}$ are the total and spin-dependent
conductivities; $\lambda_{s}$ is the spin relaxation length; $\mathcal{A}$ is the cross section of NS
boundary (of the contact). As $R_{\rm N}$, we take the total resistance of the superconducting phase
of the sample in the normal state equal to the value of the resistive jump at the superconducting
transition, assuming that the length of the spin relaxation $\lambda_{s}$ is of order of the dimension
of this phase, \emph{L}, and the whole area in the normal state is completely spin-polarized. Then
\begin{equation}\label{6}
P = \left [\frac{\delta R_{\rm N/S}}{R_{\rm N}}/(1+\frac{\delta R_{\rm N/S}}{R_{\rm N}})\right
]^{\frac{1}{2}}.
\end{equation}

\begin{figure}[htb]
\begin{center}
\includegraphics[width = 12cm]{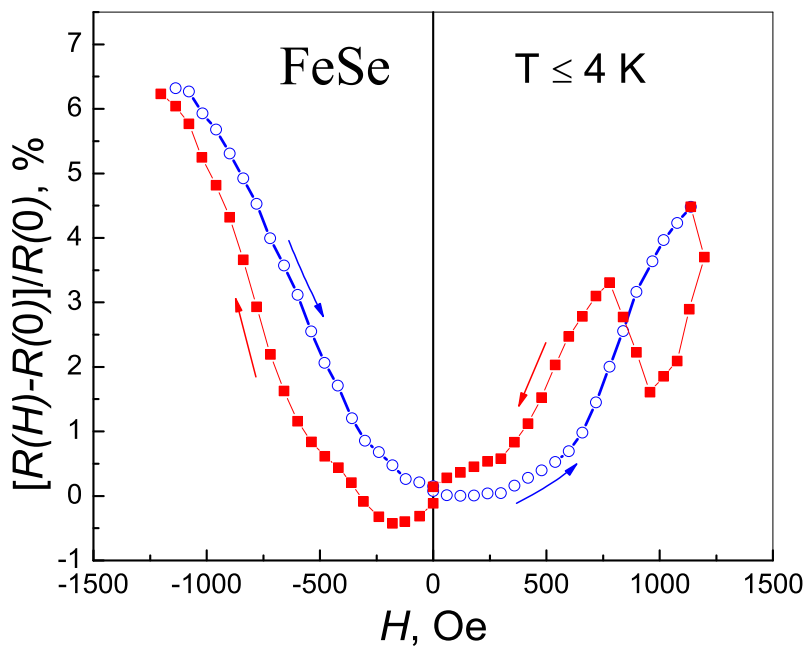}\\
\caption{Hysteresis of magnetoresistance of point-contact FeSe samples under proximity
effect.}\label{8}
\end{center}
\end{figure}

Substituting the experimental values of $\delta R_{\rm N/S}=7 \cdot 10^ {-2}\ \Omega$ (from curve 2,
Fig. 3), $\sigma^{-1} \approx 6.5 \cdot 10^ {-3}\ \Omega \cdot$cm (from independent measurement by a
standard four-probe method), $\lambda_{s} \sim L=5 \cdot 10^{-2}$ cm, and $\mathcal{A} \approx 3 \cdot
10^{-3}$ cm$^2$, we get $P \simeq 60$ \%. We shall verify that this value corresponds to that ratio of
spin-dependent conductivities at which a negative correction can exist to the conductance of
spin-polarized ferromagnetic area (F) at the transition from the F/F to F/S regime. To do this, we use
the estimates from Ref. [22] which establish the sign criterion for the resistive addition at a
similar transition. The estimates are based on the arguments that the electrons retroreflected as
Andreev holes must double their contribution to the conductance since at each reflection, a charge
2\emph{e} is transferred. Here, however, the possibility of limiting the effect is not included due to
doubling the cross section of subsequent coherent scattering by impurities in non-ballistic samples
with short mean free path, as noted above. Expressing the ratio of spin-dependent conductivities in
polarization we get this criterion in the form:
\begin{figure}[htb]
\begin{center}
\includegraphics[width = 12cm]{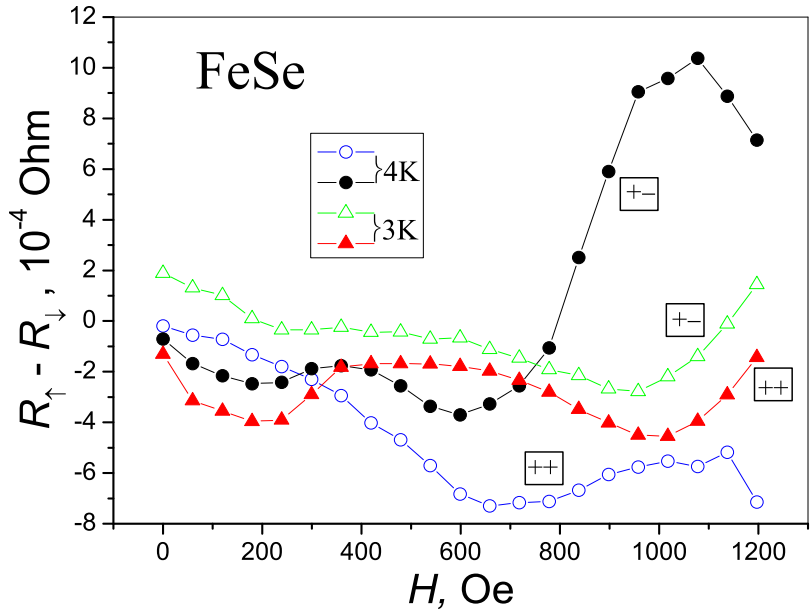}\\
\caption{Difference between the values of the magnetoresistance on the branches of the hysteresis
curves for the same values of the magnetic field and the following directions of magnetization: ++ the
same directions of the field, +— the opposite directions of the field.} \label{9}
\end{center}
\end{figure}

\begin{equation}\label{7}
\frac{\sigma_{\uparrow}}{\sigma_{\downarrow}} = \frac{1+P}{1-P} \left \{
\begin{aligned}
> 3,\ G_{\rm F/S}<G_{\rm FF(N)}\\
<3,\ G_{\rm F/S}>G_{\rm FF(N)}.
\end{aligned}
\right.
\end{equation}

Our experimental situation is entirely equivalent to the first case:
$\frac{\sigma_{\uparrow}}{\sigma_{\downarrow}} \approx 4$.

The estimate of the polarization for granular material LaO(F)FeAs by the same method (from curves 1
and 4, Fig. 4) gives a value of $P \approx 15$ \% which is in contradiction with the criterion (7).
This again suggests that the physics of conductivity of granular compounds is much more complicated
than that of single-crystal ones, in particular, because of the network of intergranular connections.

In addition to probable itinerant magnetism, we have tested the concept of spin-polarized conductivity
of FeSe for the possibility that an alternative type of magnetism exists which is characterized by
residual magnetization at reverse magnetizing. To do this, we measured the conductance of the normal
part of FeSe within boundaries defined, as a reasonable belief, by the proximity effect, in an
external magnetic field, a lot of smaller values in comparison with those converting the order
parameter to unity. The measurements show that the hysteresis of magnetoresistance does not contain an
irreversible effect at \emph{H} = 0 for any sequence of magnetizing, as can be seen from the form of
the magnetoresistance hysteresis curves shown in Fig. 8. Besides, other characteristic features of
hysteresis strike the eye (see Fig. 9): (\emph{a}) The branches belonging to different magnetic field
directions are asymmetric and (\emph{b}) the vertical displacement of the hysteresis branches in the
fields exceeding the first critical field $\simeq 30$ Oe [21] is of the same order of magnitude as the
gap features in Fig. 7. These features of the hysteresis are, of course, directly related to the spin
polarization of the current, and the absence of residual magnetoresistance at $H = 0$ for all
sequences of magnetizing shows that the nature of superconductivity in layered superconductors is
associated with the magnetism of conduction electrons (itinerant magnetism) which does not interfere
with the local magnetism of magnetic ions (such as iron ion). Note that the hysteresis of
magnetoresistance was usually observed in granular systems, but in contrast to our results obtained in
a single-crystal FeSe, it contained non-reversible values of the resistance during reversible
magnetizing. This gave rise to interpret its nature in terms of percolation mechanism of current flow
along the network of Josephson junctions [27]. The data from our single-crystal samples of FeSe
rejecting such a possibility are closest to a vortex scenario of hysteresis which is based on the
dissipative mechanism of synchronizing the vortices in the presence of defects [28].
\section{Conclusion}
We have investigated the electron transport through a barrier-free NS boundary set by the proximity
effect and the transport current inside unconventional iron-based superconductors, single-crystal
chalcogenide FeSe and granular oxypnictide LaO(F)FeAs, as parts of heterocontact samples of mesoscopic
scale.

The evidence for the spin polarization of electron transport is obtained based on the sensitivity of
Andreev conductance to symmetry with respect to the spin rotation.

The nature of the hysteresis of magnetoresistance observed in the fields much less in value than that
of the second critical field, under dispersion of the order parameter, also points to the spin
polarization of the charge carriers.

The results suggest that the nature of superconductivity in layered superconductors is connected with
the magnetism of conduction electrons (itinerant magnetism) which has nothing in common with the local
magnetism of iron ions.

\end{document}